\documentclass[12pt]{article}
\usepackage{fullpage}
\usepackage{graphicx}
\usepackage{latexsym,amsmath,amssymb,amsthm,epic,eepic,multirow}
\usepackage{natbib}

\usepackage{graphicx}
\usepackage{multirow}
\usepackage{subfigure}
\usepackage{natbib}
\usepackage{algorithm}
\usepackage{algorithmic}
\usepackage{color}

\newcommand{\EE}{\mathbb{E}}

\newcommand{\eps}{\varepsilon}

\newcommand{\argmin}{\mathrm{argmin}}

\newtheorem{theo}{Theorem}

\newtheorem{fact}{Fact}

\title{Magging: maximin aggregation\\
for inhomogeneous large-scale data}   
  
\author{Peter B\"uhlmann and Nicolai Meinshausen\\
Seminar f\"ur Statistik, ETH Z\"urich}

\begin{document}

\maketitle

\begin{abstract}
Large-scale data analysis poses both statistical and computational 
problems which need to be addressed simultaneously. 
A solution 
is often straightforward if the data are 
homogeneous: one can use classical ideas of subsampling and mean
aggregation to get a
computationally efficient solution with acceptable statistical
accuracy, where the aggregation step simply averages the results
obtained on distinct subsets of the data. However, if the data exhibit
inhomogeneities (and typically they do),
the same approach will be inadequate, as it will be unduly influenced by effects
that are not persistent across all the data due to, for example, outliers or
time-varying effects. 
We show
that a tweak to the aggregation step can produce an estimator of
effects which are common to all data, and hence interesting for
interpretation and often leading to better prediction than pooled effects. 
\end{abstract}

\section{Introduction}\label{sec.intro}
`Big data' often refers to a large collection of observations and the
associated computational issues in processing the data. 
Some of the new challenges from a statistical
perspective include:
\begin{enumerate}
\item The analysis has to be computationally efficient while retaining
  statistical efficiency \citep[cf.]{chandrasekaran2013computational}. 
\item The data are `dirty': they contain outliers, shifting
  distributions, unbalanced designs, to mention a few.
\end{enumerate}
There is also often the problem of dealing with data in real-time, which we add
to the (broadly interpreted) first challenge of computational efficiency
\citep[cf.]{MAL-035}. 

We believe that many large-scale data are inherently inhomogeneous: that
is, they are neither i.i.d. nor stationary observations from a distribution.
 Standard statistical models (e.g. linear or 
generalized linear models for regression or classification, Gaussian
graphical models) fail to capture the inhomogeneity structure in the
data. By ignoring it, prediction performance can become very poor and
interpretation of model parameters might be completely wrong. Statistical
approaches for dealing with inhomogeneous data include mixed effect models
\citep{pinheiro2000mixed}, mixture models \citep{mclachlan2004finite} and
clusterwise regression models 
\citep{desarbo1988maximum}: while they are certainly valuable in their own
right, they are typically computationally very cumbersome for large-scale
data. We present 
here a framework and methodology which addresses the issue of inhomogeneous
data while still being vastly more efficient to compute than fitting much
more complicated models such as the ones mentioned above.  

\paragraph{Subsampling and aggregation.} 
If we ignore the inhomogeneous part of the data for a moment, a simple approach to address the computational burden with large-scale data
is based on (random) subsampling: construct groups ${\cal G}_1,\ldots
,{\cal G}_G$ with ${\cal G}_g \subset \{1,\ldots ,n\}$, where $n$ denotes
the sample size and $\{1,\ldots ,n\}$ is the index set for the samples. 
The groups might be  overlapping (i.e., ${\cal G}_g \cap {\cal G}_{g'} \neq
\emptyset$ for $g \neq g'$) and do not necessarily cover the index space of
samples $\{1,\ldots ,n\}$. For every group ${\cal G}_g$, we compute an estimator
(the output of an algorithm) $\hat{\theta}_g$ and these estimates are then
aggregated to a single ``overall'' estimate
$\hat{\theta}_{\mathrm{aggr}}$, which can be achieved in different
ways. 

If we divide the data into $G$ groups of approximately equal size and
the computational complexity of the estimator scales for $n$
samples like $n^\alpha$
for some $\alpha>1$, then the subsampling-based approach above will
typically yield a computational complexity which is a factor  $G^{\alpha-1}$
faster than computing the estimator on all data, while often just
incurring an insubstantial increase in statistical error. In addition, and
importantly, effective parallel distributed computing is very easy to do
and such subsampling-based algorithms are well-suited for computation
with large-scale data.  
 
Subsampling and aggregation can thus partially address
the first challenge about feasible computation but fails for the second
challenge about proper estimation and inference in presence of
inhomogeneous data. We 
will show that a tweak to the aggregation step, which we call ``maximin
aggregation'', can often deal also with the second challenge by focusing
on effects that are common to all data (and not just mere outliers or
time-varying effects).  

\paragraph{Bagging: aggregation by averaging.}
In the context of homogeneous data, \citet{brei96} showed good
prediction performance in connection with mean or majority voting
aggregation and tree algorithms for regression or classification,
respectively. Bagging simply averages the individual estimators or
predictions.

\paragraph{Stacking and convex aggregation.}
Again in the context of homogeneous data, the following approaches have
been advocated. Instead of assigning a uniform weight to each individual
estimator as in Bagging, \citet{wolpert1992stacked} and
\citet{breiman1996stacked} proposed to learn the 
optimal weights by optimizing on a new set of data. Convex aggregation for
regression has been studied in \citet{bunea06agr}
and has been proved to lead to to approximately equally good performance as
the best member of the initial ensemble of estimators. But in fact, in
practice,  Bagging and stacking can exceed the best single estimator in
the ensemble if the data are homogeneous.  

\paragraph{Magging:  convex maximin aggregation.}
With inhomogeneous data, and in contrast to data being i.i.d.\ or
stationary realizations from a distribution, the above schemes can be
misleading as they give all data-points equal weight and can easily be misled
by strong effects which are  
present in only small parts of the data and absent for all other data. 
We show that a different type of aggregation can still lead to
consistent estimation of the effects which are common in all heterogeneous
data, the so-called maximin effects \citep{mebu14}. The maximin
aggregation, which we call Magging, is very simple and general and can
easily be implemented for large-scale data.

\section{Aggregation for regression estimators}

We now give some more details for the various aggregation schemes in the
context of linear regression models with an $n\times p$ predictor (design)
matrix $X$, whose rows correspond to $n$ samples of the $p$-dimensional
predictor variable, and with the $n$-dimensional response vector $Y\in
\mathbb{R}^n$; at this point, we do not assume a true p-dimensional
regression parameter, see also the model in (\ref{mod1}). Suppose we have
an ensemble   
of regression coefficient estimates $\hat{\theta}_g\in \mathbb{R}^p\
(g=1,\ldots,G)$, where each estimate has been obtained from the data in group
${\cal G}_g$, possibly in a  
computationally distributed fashion. The goal is to
aggregate these estimators into a single estimator
$\hat{\theta}_{\mathrm{aggr}}$.

\subsection{Mean aggregation and Bagging}
Bagging \citep{brei96} simply averages the ensemble members with equal
weight to get the
aggregated estimator 
\begin{align*} 
 \mbox{\bf Mean aggregation:}\quad \hat{\theta}_{\mathrm{aggr}} &:= \sum_{g=1}^G w_g \hat{\theta}_g,\nonumber \\
&\mbox{where } w_g=\frac{1}{G} \mbox{   for all } g=1,\ldots,G. 
\end{align*}
One could equally average the predictions $X\hat{\theta}_g$ to obtain
the predictions $X\hat{\theta}_{\mathrm{aggr}}$. The advantage of
Bagging is the simplicity of the procedure, its variance reduction
property \citep{buhlmann2002ab}, and the fact that it is not making
use of the data, which allows simple evaluation of its performance.
The term ``Bagging'' stands for \textbf{B}ootstrap
\textbf{agg}regat\textbf{ing} (mean aggregation) where the ensemble
members $\hat{\theta}_g$ are fitted on bootstrap samples of the data, that
is, the groups ${\cal G}_g$ are sampled with replacement from the whole
data.

\subsection{Stacking}
\citet{wolpert1992stacked} and \citet{breiman1996stacked} propose the idea of ``stacking'' estimators. The general idea is in
our context as follows. Let $\hat{Y}(g)= X\hat{\theta}_g\in \mathbb{R}^n$ be the
prediction of the $g$-th member in the ensemble. 
Then the stacked estimator is found as 

 \begin{align*} 
 \mbox{\bf Stacked aggregation:}\quad  \hat{\theta}_{\mathrm{aggr}} &:= \sum_{g=1}^G w_g \hat{\theta}_g,\nonumber\\
 \mbox{where } w &:= \mbox{argmin}_{w\in W} \|Y-\sum_g \hat{Y}(g) w_g\|_2,
 \end{align*}
 where the space of possible weight vectors is typically of one of the
 following forms:
\begin{align*}
 \mbox{(ridge constraint)} &:  W= \{w:\|w\|_2\le s\} \mbox{ for some }s>0 \\
 \mbox{ (sign constraint)} &:   W=\{w:\min_g w_g\ge 0 \} \\
 \mbox{ (convex constraint)} &:  W=\{w:\min_g w_g\ge 0 \mbox{ and } \sum_g w_g=1\}
\end{align*}
If the ensemble of initial estimators $\hat{\theta}_g\ (g=1,\ldots,G)$ is
derived from an independent dataset, the framework of stacked regression
has also been analyzed in \citet{bunea06agr}. Typically, though, the groups on which
the ensemble members are derived use the same underlying dataset as the
aggregation. Then, the predictions $\hat{Y}(g)$ are for each sample point
$i=1,\ldots,n$ defined as being generated with $\hat{\theta}_g^{(-i)}$,
which is the same estimator as $\hat{\theta}_g$ with observation $i$ left
out of group ${\cal G}_g$ (and consequently
$\hat{\theta}_g^{(-i)}=\hat{\theta}_g$ if $i\notin {\cal G}_g$). Instead of
a leave-one-out procedure, one could also use other leave-out schemes, such as
e.g. the out-of-bag method \citep{breiman01random}. 
To this end, we just average for a given sample over all estimators that
did not use this sample 
point in their construction, effectively setting
$\hat{\theta}_g^{(-i)}\equiv 0$ if $i\in {\cal G}_g$. The idea of
``stacking'' is thus to find the optimal linear or convex combination of
all ensemble members. The optimization is $G$-dimensional and is a
quadratic programming problem with linear inequality constraints, which can
be solved efficiently with a general-purpose quadratic programming solver.  
Note that only the inner products $\hat{Y}(g)^t \hat{Y}_{g'}$ and $\hat{Y}(g)^t Y$ for $g,g'\in \{1,\ldots,G\}$ are necessary for the optimization.

Whether stacking or simple mean averaging as in Bagging provides superior
performance depends on a range of factors. Mean averaging, as in Bagging,
certainly has an advantage in terms of simplicity. Both schemes are,
however, questionable when the data are inhomogeneous. It is then not
evident why the estimators should carry equal aggregation weight (as in
Bagging) or why the 
fit should be assessed by weighing each observation identically in the
squared error loss sense (as in stacked aggregation). 

\subsection{Magging:  maximin aggregation for heterogeneous data} 
We propose here \textbf{M}aximin \textbf{agg}regat\textbf{ing}, called
Magging, for heterogeneous data: the concept of maximin estimation has been
proposed by \citet{mebu14}, and we present a connection in Section
\ref{sec.maximin}. The differences and similarities to mean and stacked
aggregation are: 
\begin{enumerate}
\item The aggregation is a weighted average of the ensemble members
  (as in both stacked aggregation and Bagging).
\item The weights are non-uniform in general (as in stacked aggregation).
\item The weights do not depend on the response $Y$ (as in Bagging).
\end{enumerate}
The last property makes the scheme almost as simple as mean aggregation as we
do not have to develop elaborate leave-out schemes for estimation (as in
e.g. stacked regression). 
Magging is choosing the weights as a convex combination to minimize the
$\ell_2$-norm of the fitted values:
 \begin{align} 
 \mbox{\bf Magging:}\quad  \hat{\theta}_{\mathrm{aggr}} &:= \sum_{g=1}^G w_g \hat{\theta}_g,\nonumber\\
 \mbox{where } w &:= \mbox{argmin}_{w\in C_G} \|\sum_g \hat{Y}(g) w_g\|_2 \label{eq:maximin},\\
\mbox{and } C_G &:= \{ w: \min_g w_g\ge 0\mbox{ and } \sum_g w_g=1\} \nonumber .
 \end{align}
If the solution is not unique, we take the solution with lowest
$\ell_2$-norm of the weight vector among all solutions. 

The optimization and computation can be implemented in a very
efficient way.
The estimators $\hat{\theta}_g$ are computed in each group of data
${\cal G}_g$ separately, and this task can be easily performed in
parallel. In the end, the estimators only need to be 
combined by calculating optimal convex weights in $G$-dimensional space
(where typically 
$G\ll n$ and $G\ll p$) with quadratic programming; some
pseudocode in \texttt{R} \citep{R14} for these convex weights is presented
in the Appendix. Computation of Magging 
is thus computationally often massively faster and simpler than a related 
direct estimation estimation scheme proposed in \citet{mebu14}. Furthermore, 
Magging is very generic (e.g. one can choose its own favored regression
estimator $\hat{\theta}_g$ for the $g$-th group) and also straightforward
to use in more general settings beyond linear models.   

The Magging scheme will be motivated in the following Section
\ref{sec.maximin} with a model for inhomogeneous data and it will be shown
that it corresponds to maximizing the minimally ``explained variance''
among all data groups. The main idea is that if an effect is common across
all groups ${\cal G}_g\ (g=1,\ldots ,G)$, then we cannot ``average it
away'' by searching for a specific convex combination of the weights. The
common effects will be present in all groups and will thus be retained even
after the minimization of the aggregation scheme. 

The construction of the groups ${\cal G}_g\ (g=1,\ldots ,G)$ for
Magging in presence of inhomogeneous data is rather specific and described
in Section \ref{subsec.genunknown} for various scenarios. There, 
Examples 1 and 2 represent the setting where the data within
each group is (approximately) homogeneous, whereas Example 3
is a case with randomly subsampled groups, despite the fact of
inhomogeneity in the data. 

\section{Inhomogeneous data and maximin effects}\label{sec.maximin}
 
We motivate in the following why Magging (maximin aggregation) can be
useful for inhomogeneous data when the interest is on effects that are
present in all groups of data.  

In the linear model setting, we consider the framework of 
a mixture model 
\begin{eqnarray}\label{mod1}
Y_i = X_i^t B_i + \eps_i,\ i=1,\ldots ,n,
\end{eqnarray}
where $Y_i$ is a univariate response variable, $X_i$ is a $p$-dimensional
covariable, $B_i$ is a $p$-dimensional regression parameter, and $\eps_i$
is a stochastic noise term with mean zero and which is independent of the
(fixed or random) covariable. Every sample point $i$ is allowed to have its
own and different regression parameter: hence, 
the inhomogeneity occurs because of changing parameter vectors, and 
we have 
a mixture model where, in principle, every sample arises from a
different mixture component. The model in (\ref{mod1}) is often too
general: we make the assumption that the regression parameters $B_1,\ldots
,B_n$ are realizations from a distribution $F_B$:
\begin{eqnarray}\label{mod2}
B_i \sim F_B,\ i=1,\ldots ,n,
\end{eqnarray}
where the $B_i$'s do not need to be independent of each other. However, we
assume that the $B_i$'s are independent from the $X_i$'s and $\eps_i$'s. 

\medskip
\emph{Example 1: known groups.} Consider the case where there are known
groups ${\cal G}_g$ with $B_i \equiv  b_g$ for all $i \in {\cal
  G}_g$. Thus, this is a clusterwise regression problem (with
\emph{known} clusters) where every group ${\cal G}_g$ has the same
(unknown) regression parameter vector $b_g$. We note that the groups ${\cal
  G}_g$ are the ones for constructing the Magging estimator described in
the previous section. 

\medskip
\emph{Example 2: smoothness structure.} Consider the
situation where there is a smoothly changing behavior of the $B_i$'s with
respect to the sample indices $i$: this can be achieved by positive
correlation among the $B_i$'s. In practice, the sample index
often corresponds to time. There are no true (unknown) groups in this
setting. 

\medskip
\emph{Example 3: unknown groups.} This is the same setting as in Example 1
but the groups ${\cal G}_g$ are unknown. From an estimation point of view,
there is a substantial difference to Example~1 \citep{mebu14}.   
\medskip

\subsection{Maximin effects}

In model (\ref{mod1}) and in the Examples 1--3 mentioned above, we have a
``multitude'' of regression parameters. We aim for a single $p$-dimensional
parameter, which contains the common components among all $B_i$'s (and
essentially sets the non-common components to the value zero). This can be
done by the idea of so-called maximin effects which we explain next. 

Consider a linear model with the fixed $p$-dimensional regression parameter
$b$ which can take values in the support of $F_B$ from (\ref{mod2}):
\begin{eqnarray}\label{mod.b}
Y_i = X_i^t b + \eps_i,\ i=1,\ldots ,n,
\end{eqnarray}
where $X_i$ and $\eps_i$ are as in (\ref{mod1}) and assumed to be i.i.d. We
will connect the random 
variables $B_i$ in (\ref{mod1}) to the values $b$ via a worst-case
analysis as described below: for that purpose, the parameter $b$ is assumed
to not depend on the sample index $i$. The variance 
which is explained by choosing a parameter vector $\beta$ in the linear
model (\ref{mod.b}) is  
\begin{eqnarray*}
V_{\beta,b} := \EE|Y|^2 - \EE|Y - X^t \beta|^2 = 2 \beta^t \Sigma b -
\beta^t \Sigma \beta,
\end{eqnarray*}
where $\Sigma$ denotes the covariance matrix of $X$. We aim for maximizing
the explained variance in the worst (most adversarial) scenario: this is
the definition of the maximin effects. 

\medskip
\emph{Definition \citep{mebu14}.} The maximin effects parameter is
\begin{eqnarray*}
b_{\mathrm{maximin}} = \argmin_{\beta} \max_{b \in \mathrm{supp}(F_B)} -
V_{\beta,b},
\end{eqnarray*}
and note that the definition uses the negative explained variance $-V_{\beta,b}$. 

\begin{figure}[!h]
\begin{center}
\includegraphics[scale=0.35]{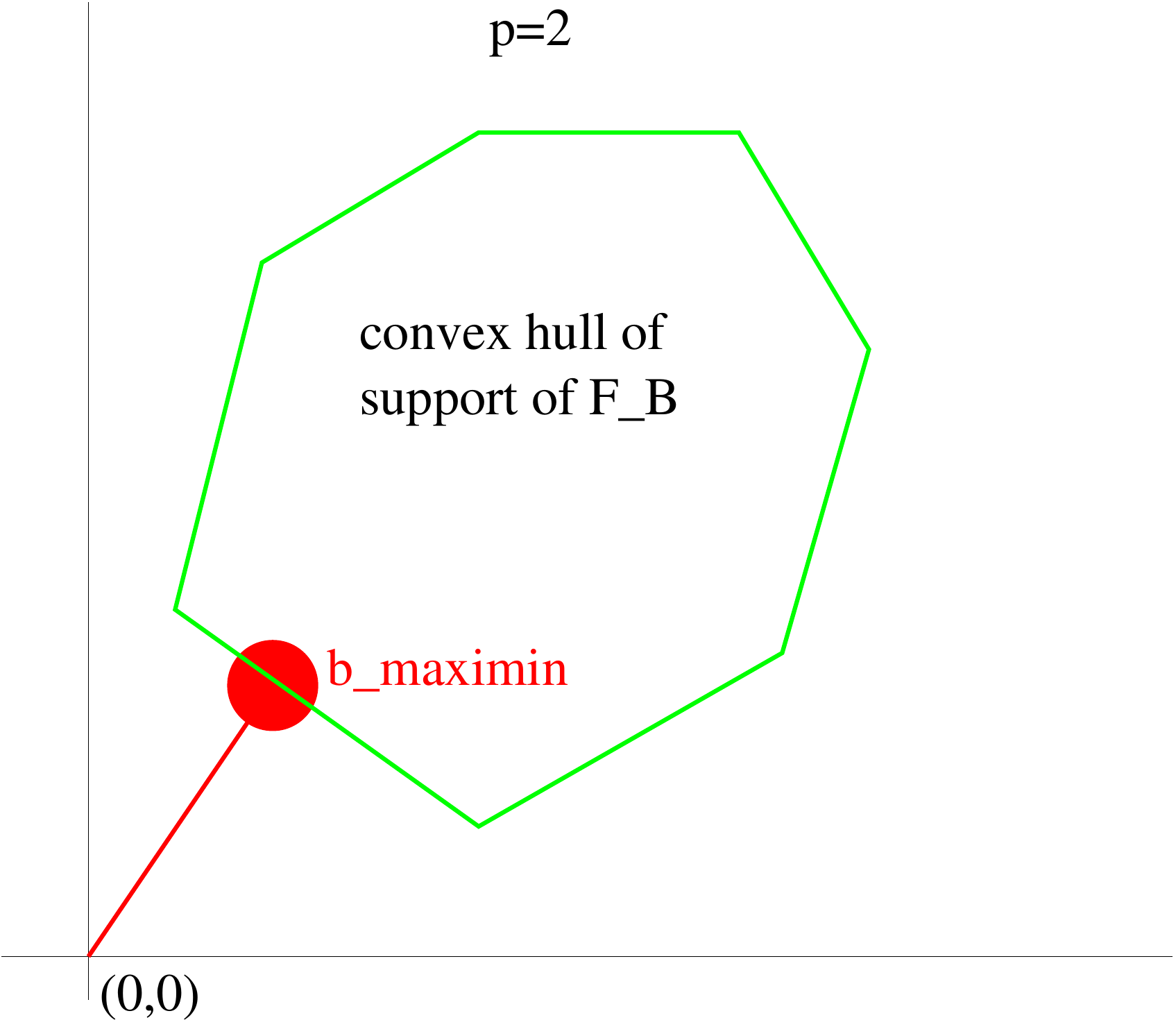}
\end{center}
\caption{Illustration of Fact \ref{fact1} in dimension
  $p=2$.}\label{fig.charact} 
\end{figure}

\medskip
The maximin effects can be interpreted as an aggregation among the support
points of $F_B$ to a single parameter vector, i.e., among all the $B_i$'s
(e.g. in Example~2) or among all 
the clustered values $b_g$ (e.g. in Examples 1 and 3), see also Fact
\ref{fact1} below. The maximin effects
parameter is different from the pooled effects
$b_{\mathrm{pool}} = \argmin_{\beta} \; \EE_B[-V_{\beta,B}]$ and a bit
surprisingly, also rather different from the prediction analogue
\begin{eqnarray*}
b_{\mathrm{pred-maximin}} = \argmin_{\beta} \max_{b \in \mathrm{supp}(F_B)}
\EE[(X^t b - X^t \beta)^2]. 
\end{eqnarray*}
In particular, the value zero has a special status for the maximin effects
parameter $b_{\mathrm{maximin}}$, unlike for $b_{\mathrm{pred-maximin}}$ or
$b_{\mathrm{pool}}$, see \citet{mebu14}. The following is an important ``geometric''
characterization which indicates the special status of the value zero, see
also Figure \ref{fig.charact}.  
\begin{fact}\label{fact1}\citep{mebu14}
Let $H$ be the convex hull of the support of $F_B$. Then
\begin{eqnarray*}
b_{\mathrm{maximin}} = \argmin_{\gamma \in H} \; \gamma^t \Sigma \gamma. 
\end{eqnarray*}
That is, the maximin effects parameter $b_{\mathrm{maximin}}$ is the point in the
convex hull $H$ which is closest to zero with respect to the distance
$d(u,v) = (u-v)^t \Sigma (u-v)$: in particular, if the value zero is in
$H$, the maximin effects parameter equals $b_{\mathrm{maximin}} \equiv 0$. 
\end{fact}
The characterization in Fact \ref{fact1} leads to an interesting robustness
issue which we will discuss below in Section \ref{subsec.robustness}. 

The connection to Magging (maximin aggregation) can be made most easily
for the setting of Example 1 with known groups and constant regression
parameter $b_g$ within each group ${\cal G}_g$. We can rewrite, using Fact
\ref{fact1}: 
\begin{eqnarray*}
& &b_{\mathrm{maximin}} = \sum_{g=1}^G w_g^0 b_g,\\
& &w^0 = (w_1^0,\ldots ,w_G^0) = \argmin_{w \in C_G} \sum_{g,g'=1}^G w_g w_{g'}
b_g^T \Sigma b_g = \argmin_{w \in C_G} \EE_X\|\sum_{g=1}^G w_g X b_g\|_2^2,
\end{eqnarray*}
where $C_G$ is as in (\ref{eq:maximin}). The Magging estimator is then
using the plug-in principle with estimates $\hat{\theta}_g$ for $b_g$ and
$\|\sum_g w_g \hat{Y}(g)\|_2^2$ for $\EE_X\|\sum_{g=1}^G w_g X b_g\|_2^2$. 

\subsection{Robustness}\label{subsec.robustness}

\begin{figure}[!htb]
\begin{center}
\includegraphics[scale=0.25]{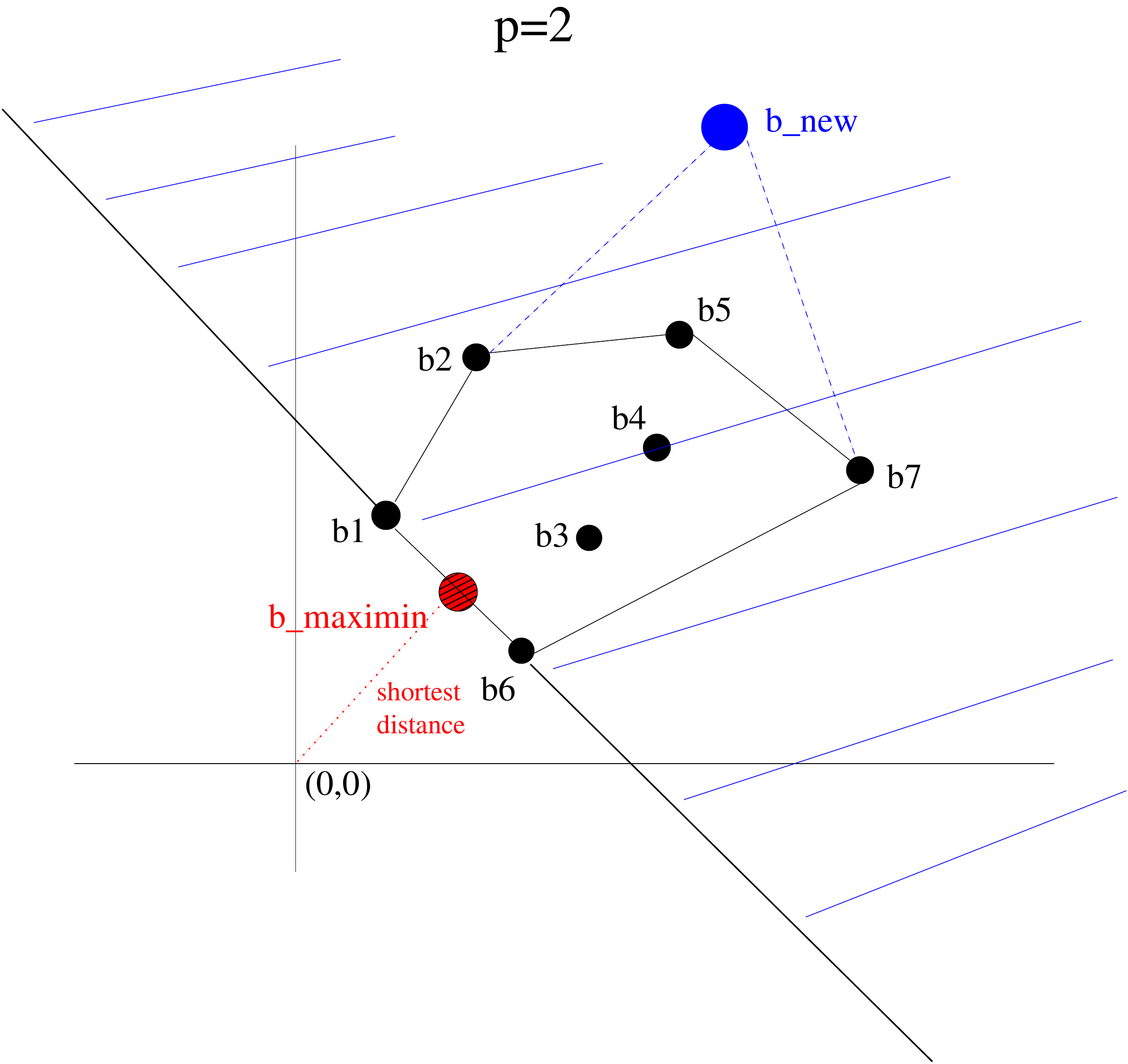}
\hspace*{15mm}
\includegraphics[scale=0.25]{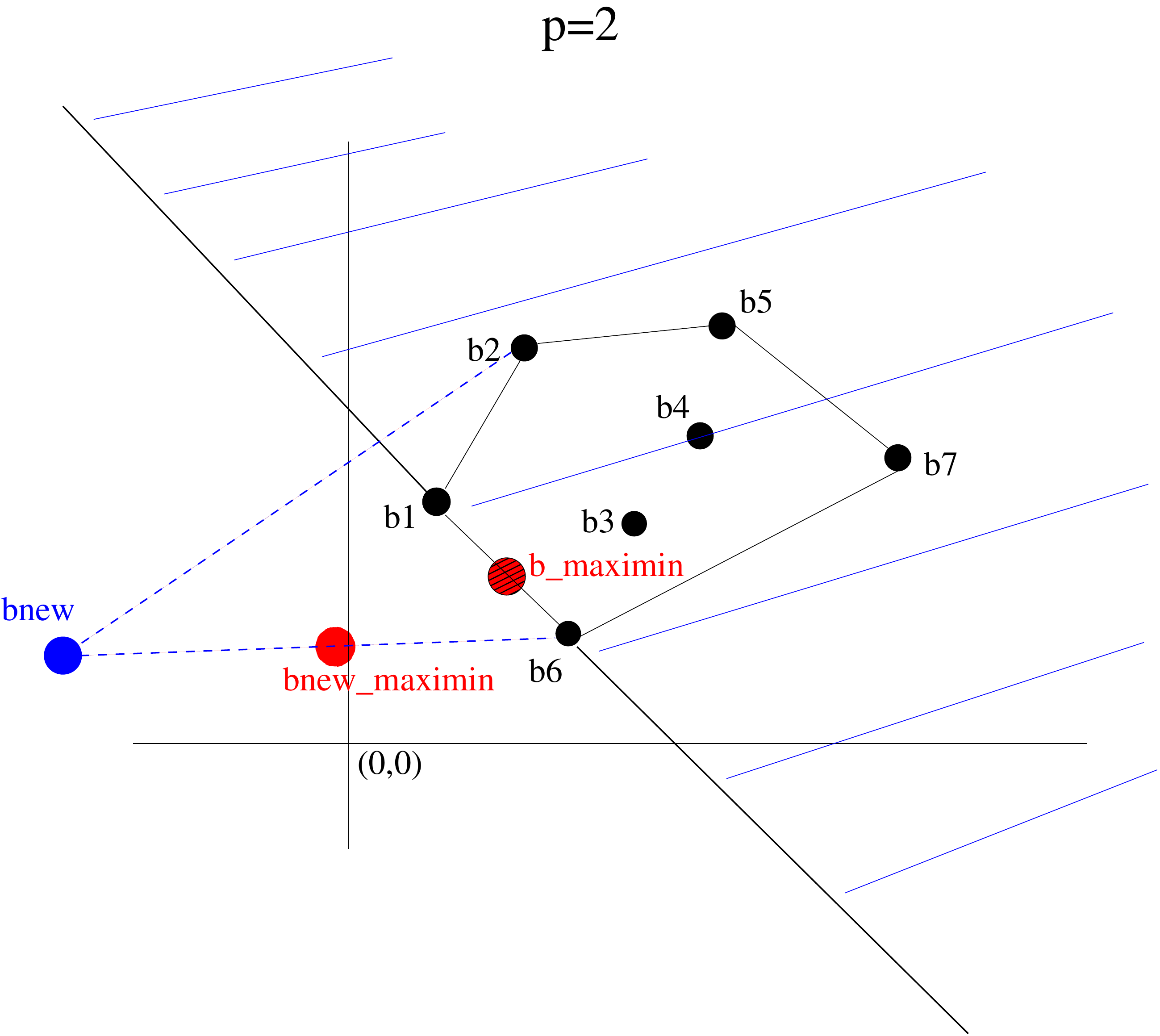}
\end{center}
\caption{Illustration of the case with a finite number of possible values
  for $B$.
Left panel: The values $b_1,\ldots ,b_7$   are
  possible realizations of $B_i$, and 
  $b_{\mathrm{maximin}}$ is the closest point to zero in the convex hull of
  $\{b_1,\ldots ,b_7\}$ (in black). When adding a new
  additional realization 
  $b_{\mathrm{new}}$, the convex hull becomes larger (in dashed
  blue). As long as the new  support point is in the blue shaded half-space,
  the maximin effects parameter $b_{\mathrm{maximin}}$ remains the same
  \emph{regardless} of how far away the new support point is added. Right
  panel: A new
  additional realization 
  $b_{\mathrm{new}}$ arises which does not lie in the blue shaded
  half-space, the convex hull becomes larger (in dashed blue) and 
  the new maximin effects parameter becomes
  $b_{\mathrm{new,maximin}}$. Since the new convex hull (in dashed blue)
  gets enlarged by a new realized value $b_{\mathrm{new}}$ , the corresponding
  new maximin effects parameter $b_{\mathrm{new,maximin}}$ must be closer
  to the origin than the original  parameter $b_{\mathrm{maximin}}$. Thus, it is
  impossible to shift $b_{\mathrm{maximin}}$ away from zero by placing new
 realizations at arbitrary positions.}
\label{fig.outlier1}
\end{figure}
%

It is instructive to see how the maximin effects parameter is changing if
the support of $F_B$ is extended, possibly rendering the support non-finite.
There are two possibilities, illustrated by Figure~\ref{fig.outlier1}.
In the first case, illustrated in the left panel of
Figure~\ref{fig.outlier1}, the new parameter vector 
$b_{\mathrm{new}}$ is not changing the point in the convex hull of the
support of $F_B$ that is 
closest to the origin. The maximin effects parameter is then unchanged. The
second 
situation is illustrated in the right panel of Figure~\ref{fig.outlier1}. The addition of a
new support point here does change the convex hull of the support such that
there is now a point in the support closer to the origin. Consequently, the
maximin effects parameter will shift to this new value.  The maximin
effects parameter thus 
is either unchanged or is moving closer to the origin. Therefore, maximin
effects parameters and their 
estimation exhibit an excellent robustness feature with respect to breakdown properties. 

\subsection{Statistical properties of Magging}\label{subsec.Maggingprop}
%
We will derive now some statistical properties of Magging, the maximin
aggregation scheme, proposed in (\ref{eq:maximin}). They depend also on the
setting-specific construction of the groups ${\cal G}_1,\ldots {\cal G}_G$
which is described in Section \ref{subsec.genunknown}. 

\paragraph{Assumptions.} 
Consider the model (\ref{mod1}) and that there are $G$ groups ${\cal G}_g\
(g=1,\ldots,G)$ of data 
samples. Denote by $Y_g$ and $X_g$ the
data values corresponding to group ${\cal G}_g$.
\begin{description}
\item[(A1)] Let $b^*_g$ be the optimal regression vector in each
  group, that is $b_g^* = \EE_B[|{\cal G}_g|^{-1} \sum_{i \in {\cal G}_g}
  B_i]$. Assume that $b_{\mathrm{maximin}}$ is in the convex hull of $\{b^*_1,\ldots,b^*_G\}$.
\item[(A2)] We assume random design with a mean-zero random predictor
  variable $X$ with covariance matrix $\Sigma$ and let
  $\hat{\Sigma}_g=|{\cal G}_g|^{-1} X_g^t X_g$ be the empirical Gram
  matrices. Let $\hat{\theta}_g\ (g=1,\ldots,G)$ be the estimates in each
  group. Assume that there exists some $\eta_1,\eta_2>0$ such that 
\begin{align*}
\max_g (\hat{\theta}_g -b^*_g)^t \Sigma (\hat{\theta}_g -b^*_g) & \le
\eta_1,\nonumber \\ 
\max_g \|\hat{\Sigma}_g - \Sigma\|_\infty & \le \eta_2 ,
\end{align*}
where $m=\min_g |{\cal G}_g|$ is the minimal sample size across all
groups.
\item[(A3)] The optimal and estimated vectors are sparse in the sense that there exists some $\kappa>0$ such that
\begin{equation*}
\max_g \|b^*_g\|_1 \le \kappa \quad\mbox{ and } \quad \max_g \|\hat{\theta}_g\|_1 \le \kappa.
\end{equation*}
\end{description}

Assumption (A1) is fulfilled for known groups, where the convex hull
of $\{b^*_1,\ldots,b^*_G\}$ is equal to the convex hull of the support
of $F_B$ and the maximin-vector $b_{\mathrm{maximin}}$ is hence contained
in the former. Example 1 is fulfilling the requirement, and we 
will discuss generalizations to the settings in Examples 2 and 3 below in
Section \ref{subsec.genunknown}. Assumptions (A2) and (A3) are relatively
mild: the first part of (A3) is an assumption that the underlying model is
sufficiently sparse.
If we consider standard Lasso estimation with sparse optimal coefficient
vectors and assuming bounded predictor variables, then (A2) is fulfilled 
with high probability for $\eta_1$ of the order
  $\kappa(\log(pG)/m)^{1/2}$ (faster rates are possible under
 a compatibility assumption) and $\eta_2$ of order $\log(pG)/m$,
where  $m=\min_g|{\cal G}_g|$ denotes the minimal sample size across all groups; see for
see for example~\citet{mebu14}. 

Define for $x\in \mathbb{R}^p$, the norm $\|x\|_\Sigma^2 = x^t\Sigma x$ and
let $\hat{\theta}_{\mathrm{Magging}}$ be the Magging estimator
(\ref{eq:maximin}).
\begin{theo}\label{theo:main}  Assume (A1)-(A3). Then 
\[ \|\hat{\theta}_{\mathrm{Magging}}-b_{\mathrm{maximin}}\|_\Sigma^2  \;\le\;  6\eta_1 + 4\eta_2 \kappa^2  .\]
\end{theo}
A proof is given in the Appendix. 

The result implies that the maximin effects parameter can be estimated with good
accuracy by Magging (maximin aggregation) if the individual effects in each
group can be estimated accurately with standard methodology (e.g. penalized
regression methods). 


\subsubsection{Construction of groups and their validity for different settings}\label{subsec.genunknown}

Theorem \ref{theo:main} hinges mainly on assumption (A1). We discuss
the validity of the assumption for the three discussed settings
 under appropriate (and setting-specific)
sampling of the data-groups.

\medskip
\emph{Example 1: known groups (continued).}
Obviously, the groups ${\cal G}_g\ (g=1,\ldots,G)$ are chosen to be the true known
groups. 

Assumption (A1) is then trivially fulfilled with known groups and constant
regression parameter within groups (clusterwise regression). 

\medskip
\emph{Example 2: smoothness structure (continued).}
We construct $G$ groups
of non-overlapping consecutive observations. For simplicity, we would
typically use equal group size $m = \lfloor n/G \rfloor$ so that ${\cal
  G}_1 = \{1,2,\ldots ,m\}, {\cal G}_2 = \{m+1,\ldots ,2m\},\ldots ,{\cal G}_G =
\{(G-1)m+1,\ldots ,n\}$.

When
taking sufficiently  
many groups and for a certain model of smoothness structure, condition~(A1)
will be fulfilled with high probability \citep{mebu14}: it is shown there
that it is rather likely to get some groups of consecutive observations 
where the optimal vector is approximately constant and the convex hull of these
``pure'' groups will be equal to the convex hull of the support of $F_B$.

\medskip
\emph{Example 3: unknown groups (continued).} We construct $G$ groups of
equal size $m$ by random subsampling: sample without replacement within a
group and with replacement between groups.  

This random subsampling
strategy can be shown to fulfill condition~(A1)
when assuming an additional so-called Pareto condition \citep{mebu14}. As an
example, a model with a fraction of outliers fulfills (A1) and one obtains an
important robustness property of Magging which is closely connected to
Section \ref{subsec.robustness}. 

\subsection{Numerical example}

\begin{figure}
\begin{center}
\includegraphics[width=0.99\textwidth]{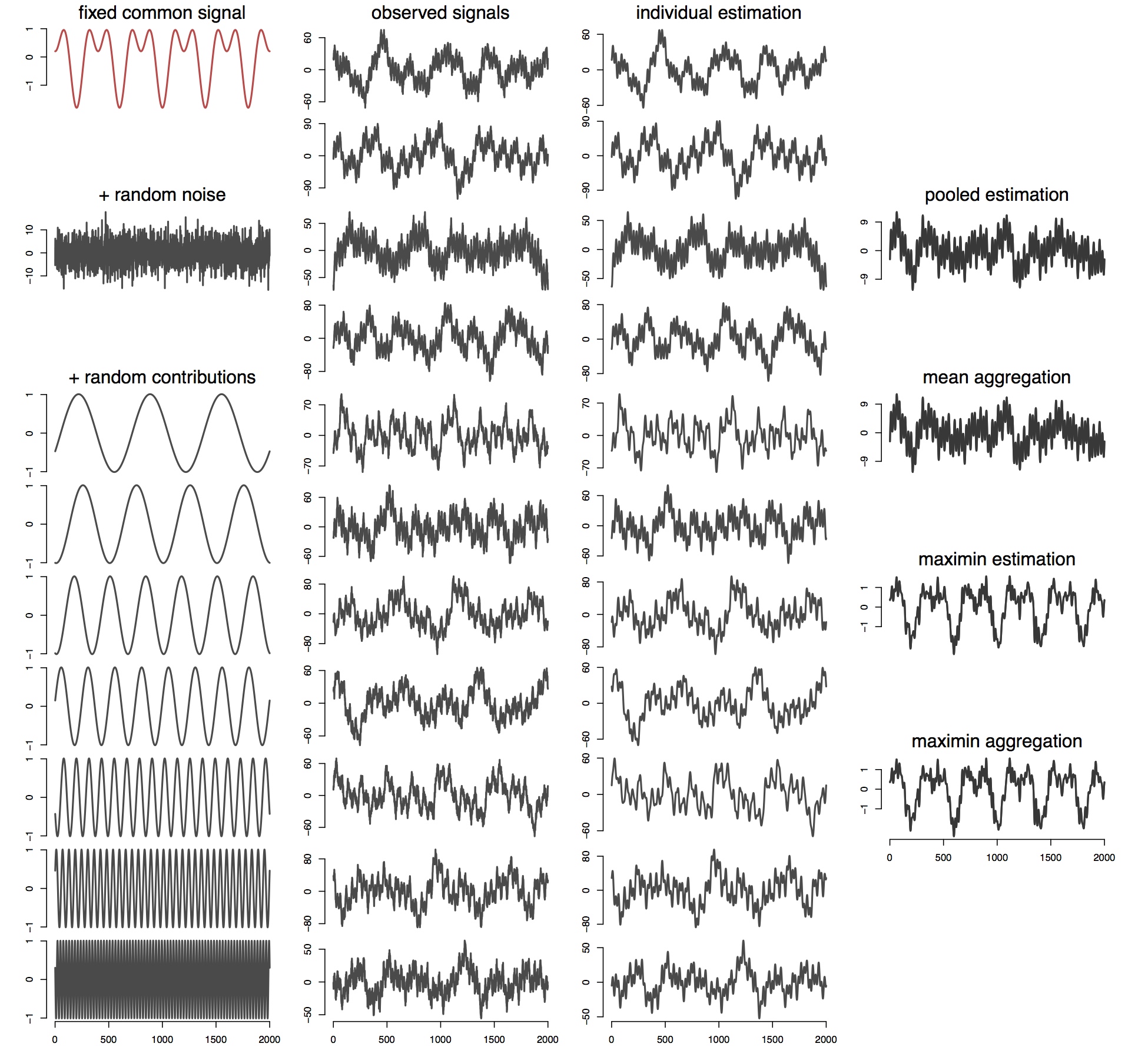}
\end{center}
\caption{The left column shows the data generation. Each group has the same
  fixed common effect (shown in red at the top left), and gets random noise
  as well as other random periodic contributions added (with random phase),
  where the latter two contributions are drawn independently for all groups
  $g=1,\ldots,G=50$. The second column shows the realizations of $Y_g$ for
  the first groups $g=1,\ldots,11$, while the third shows the least-squares
  estimates of the signal when projecting onto the space of periodic
  signals in a certain frequency-range. The last column shows from top to
  bottom: (a) the pooled estimate one obtains when adding all groups into
  one large dataset and estimating the signal on all data simultaneously
  (the estimate does not match closely the common effects shown in red);
  (b) the mean aggregated data obtained by averaging the individual
  estimates (here identical to pooled estimation); (c) the (less generic)
  maximin effects estimator from \citet{mebu14}, and (d) Magging: maximin
  aggregated estimators~(\ref{eq:maximin}), both of which match the common
  effects quite closely.     }\label{fig.example} 
\end{figure}

We illustrate the difference between mean aggregation and maximin
aggregation (Magging) with a simple example. 
We are recording, several times, data in a time-domain. Each recording (or
group of observations) contains a common signal, a combination of two
frequency components, shown in the top left of Figure~\ref{fig.example}. On
top of the common signal, seven out of a total of 100 possible frequencies
(bottom left in Figure~\ref{fig.example}) add to the recording in each
group with a random phase. The 100 possible frequencies are the first
frequencies $2\pi  j/P$, $j=1,\ldots,100$ for periodic signal with
periodicity $P$ defined by the length of the recordings. They form the
dictionary used for estimation of the signal.  In total $G=50$ recordings
are made, of which the first 11 are shown in the second column of
Figure~\ref{fig.example}. The estimated signals are shown in the third
column, removing most of the noise but leaving the random contribution from
the non-common signal in place. Averaging over all estimates in the mean
sense yields little resemblance with the common effects. The same holds
true if we estimate the coefficients by pooling all data into a single
group (first two panels in the rightmost column of
Figure~\ref{fig.example}). 
Magging (maximin aggregation) and the closely related but less generic
maximin estimation \citep{mebu14}, on the other hand, 
approximate the common signal in all groups quite well (bottom two panels
in the rightmost column of Figure~\ref{fig.example}).  

\citet{mebu14} provide other real data results where maximin effects
estimation leads to better out-of-sample predictions in two financial
applications. 

\section{Conclusions}

Large-scale and `Big' data poses many challenges from a statistical
perspective. One of them is to develop algorithms and methods that retain
optimal or reasonably good 
statistical properties while being computationally cheap to compute. 
Another is to deal with inhomogeneous data which might contain
outliers, shifts in distributions and other effects that do not fall
into the classical framework of identically distributed or stationary
observations. Here we have 
shown how Magging (``maximin aggregation'') can be a useful approach
addressing both of the two challenges. 
The whole task is split into several smaller
datasets (groups), which can be 
processed trivially in parallel. The standard 
solution is then to average the results from all tasks, which we call
``mean aggregation'' here. In contrast, we show that finding a certain convex
combination, we can
detect the signals which are common in all subgroups of the
data. While ``mean aggregation'' is easily confused by signals that
shift over time or which are not present in all groups,
Magging (``maximin aggregation'') eliminates as much as possible these
inhomogeneous effects and just retains the common signals which is an
interesting feature in its own right and often improves out-of-sample
prediction performance.  

\bibliographystyle{apalike}

\begin{thebibliography}{}

\bibitem[Breiman, 1996a]{brei96}
Breiman, L. (1996a).
\newblock Bagging predictors.
\newblock {\em Machine Learning}, 24:123--140.

\bibitem[Breiman, 1996b]{breiman1996stacked}
Breiman, L. (1996b).
\newblock {Stacked regressions}.
\newblock {\em Machine Learning}, 24:49--64.

\bibitem[Breiman, 2001]{breiman01random}
Breiman, L. (2001).
\newblock {Random Forests}.
\newblock {\em Machine Learning}, 45:5--32.

\bibitem[B\"uhlmann and Yu, 2002]{buhlmann2002ab}
B\"uhlmann, P. and Yu, B. (2002).
\newblock Analyzing bagging.
\newblock {\em The Annals of Statistics}, 30:927--961.

\bibitem[Bunea et~al., 2007]{bunea06agr}
Bunea, B., Tsybakov, A., and Wegkamp, M. (2007).
\newblock Aggregation for {G}aussian regression.
\newblock {\em The Annals of Statistics}, 35:1674--1697.

\bibitem[Chandrasekaran and Jordan, 2013]{chandrasekaran2013computational}
Chandrasekaran, V. and Jordan, M.~I. (2013).
\newblock Computational and statistical tradeoffs via convex relaxation.
\newblock {\em Proceedings of the National Academy of Sciences},
  110:E1181--E1190.

\bibitem[DeSarbo and Cron, 1988]{desarbo1988maximum}
DeSarbo, W. and Cron, W. (1988).
\newblock A maximum likelihood methodology for clusterwise linear regression.
\newblock {\em Journal of Classification}, 5:249--282.

\bibitem[Mahoney, 2011]{MAL-035}
Mahoney, M.~W. (2011).
\newblock Randomized algorithms for matrices and data.
\newblock {\em Foundations and Trends® in Machine Learning}, 3:123--224.

\bibitem[McLachlan and Peel, 2004]{mclachlan2004finite}
McLachlan, G. and Peel, D. (2004).
\newblock {\em Finite Mixture Models}.
\newblock John Wiley \& Sons.

\bibitem[Meinshausen and B{\"u}hlmann, 2014]{mebu14}
Meinshausen, N. and B{\"u}hlmann, P. (2014).
\newblock Maximin effects in inhomogeneous large-scale data.
\newblock Preprint arXiv:1406.0596.

\bibitem[Pinheiro and Bates, 2000]{pinheiro2000mixed}
Pinheiro, J. and Bates, D. (2000).
\newblock {\em Mixed-effects Models in S and S-PLUS}.
\newblock Springer.

\bibitem[{R Core Team}, 2014]{R14}
{R Core Team} (2014).
\newblock {\em R: A Language and Environment for Statistical Computing}.
\newblock R Foundation for Statistical Computing, Vienna, Austria.

\bibitem[Wolpert, 1992]{wolpert1992stacked}
Wolpert, D. (1992).
\newblock {Stacked generalization}.
\newblock {\em Neural {N}etworks}, 5:241--259.

\end{thebibliography}

\section*{Appendix}

{\it Proof of Theorem~\ref{theo:main}:} 
Define for $w\in C_G$ (where $C_G\subset \mathbb{R}^G$ is as defined in (\ref{eq:maximin}) the set of positive vectors that sum to one),
\begin{eqnarray*}
 \hat{\theta}(w) :=  \sum_{g=1}^G w_g \hat{\theta}_g \quad\mbox{ and } \quad
\theta(w) := \sum_{g=1}^G w_g b^*_g
\end{eqnarray*}
And let for $\hat{\Sigma} = n^{-1} X^tX$,
\begin{eqnarray*}
 \hat{L}(w) :=  \hat{\theta}(w)^t \hat{\Sigma} \hat{\theta}(w)  \quad\mbox{ and }\quad 
L(w) := \theta(w) ^t \Sigma \theta(w).
\end{eqnarray*}
Then $w^* = \argmin_w L(w)$ and $b_{\mathrm{maximin}}=\theta(w^*)$ and $\hat{w} = \argmin_w \hat{L}(w)$ and $\hat{\theta}_{\mathrm{Magging}} = \hat{\theta}(\hat{w})$. Now, using (A3)
\begin{align*}
\sup_{w\in C_G} | \hat{L}(w) - L(w)| &\le \sup_{w\in C_G} |\theta(w) ^t (\Sigma-\hat{\Sigma}) \theta(w)| + \max_{g} \| b^*_g-\hat{b}_g\|_\Sigma^2  \\ & \le \eta_2 (\max_{w\in C_G} \|\theta(w)\|_1)^2  + \eta_1.
\end{align*}
Hence, as $w^*=\argmin_{w\in C_G} L(w)$ and $\hat{w} = \argmin_w \hat{L}(w)$,
\begin{equation} \label{eq:1} L(\hat{w}) \le L(w^*) + 2(\eta_1+\eta_2\kappa^2). \end{equation}
For $\Delta:=  \theta(\hat{w}) - \theta(w^*)$,
\begin{align*}
L(\hat{w}) = \| \theta(\hat{w}) \|_\Sigma^2  & = (\theta(w^*)+\Delta)^t \Sigma (\theta(w^*)+\Delta) \\
&= \theta(w^*)^t \Sigma \theta(w^*) + 2\Delta^t \Sigma \theta(w^*) + \Delta^t \Sigma \Delta \\
&\ge L(w^*) + \|\Delta\|_\Sigma^2,
\end{align*}
where $\Delta^t \Sigma \theta(w^*) \ge 0$ follows by the definition of the maximin vector $\theta(w^*)  = b_{\mathrm{maximin}}$.
Combining the last inequality with (\ref{eq:1}),
\begin{equation}\label{eq:2} \|\theta(\hat{w}) - \theta(w^*)\|_\Sigma^2 \le 2(\eta_1+\eta_2\kappa^2)\end{equation}
Furthermore, by (A3),
\[ \sup_{w\in C_G} \| \hat{\theta}(w) - \theta(w)\|_\Sigma^2 \le \eta_1 .\]
 Using the equality for $\hat{\theta}_{\mathrm{Magging}}=\hat{\theta}(\hat{w})$, 
\begin{equation}\label{eq:3} \| \hat{\theta}(\hat{w}) - \theta(\hat{w})\|_\Sigma^2 \le \eta_1 .\end{equation}
Combining (\ref{eq:2}) and (\ref{eq:3}),
\begin{align*} \| \hat{\theta}_{\mathrm{Magging}} - b_{\mathrm{maximin}}\|_\Sigma^2 = \| \hat{\theta}(\hat{w}) - 
\theta(w^*) \|_\Sigma^2 & \le 2 \big( \| \hat{\theta}(\hat{w}) - \theta(\hat{w})\|_\Sigma^2  + \|\theta(\hat{w}) - \theta(w^*)\|_\Sigma^2\big) \\ &\le 2\big( \eta_1 + 2(\eta_1+\eta_2\kappa^2)\big) \\ &= 6\eta_1+4\eta_2\kappa^2,\end{align*}
which completes the proof.\hfill$\Box$

\bigskip\noindent
\emph{Implementation of Magging in \texttt{R}:}\\
We present here some pseudo-code for computing the weights $w_1,\ldots
,w_G$ in Magging (\ref{eq:maximin}), using quadratic programming in the
\texttt{R}-software environment.
\begin{verbatim}
library(quadprog)
theta <- cbind(theta1,...,thetaG)   #matrix with G columns:
                                    #each column is a regression estimate

hatS <- t(X) %*% X/n                #empirical covariance matrix of X
H <- t(theta) %*% hatS %*% theta    #assume that it is positive definite
                                    #(use H + xi * I, xi > 0 small, otherwise)
A <- rbind(rep(1,G),diag(1,G))      #constraints
b <- c(1,rep(0,G))              
d <- rep(0,G)                       #linear term is zero
w <- solve.QP(H,d,t(A),b, meq = 1)  #quadratic programming solution to 
                                    #argmin(x^t H x) such that Ax >= b and 
                                    #first inequality is an equality  
\end{verbatim}

\end{document}